\documentclass[a4paper,11pt]{article}

\usepackage{jheppub}
\usepackage{multicol}
\usepackage{multirow}
\usepackage{subfigure}
\usepackage{slashed} 
\newcommand{\met}{\ensuremath{\slashed{E}_T}}
\newcommand{\wwa}{{\sc $W^+W^-\gamma$ }}
\newcommand{\waa}{{\sc $W\gamma\gamma$ }}
\newcommand{\wwaa}{{\sc $WW\gamma\gamma$ }}
\newcommand{\wwza}{{\sc $WWZ\gamma$ }}

\newcommand{\fbinv} {\mbox{\ensuremath{\,\text{fb}^\text{$-$1}}}}

\newcommand{\pythia}{{\sc Pythia}}
\newcommand{\delphes}{{\sc Delphes}}

\newcommand{\mgme}{{\sc MadGraph/MadEvent}}

\newcommand{\madgraph}{{\sc MadGraph}}
\newcommand{\tauola}{{\sc TAUOLA}}
\newcommand{\madevent}{{\sc MadEvent}}

\title{Probing $W^+W^-\gamma$ Production and Anomalous Quartic Gauge Boson Couplings at the CERN LHC}

\author{Daneng Yang,}
\author{Yajun Mao,}
\author{Qiang Li,}
\author{Shuai Liu,}
\author{Zijun Xu,}
\author{Ke Ye}

\affiliation{Department of Physics and State Key Laboratory of Nuclear Physics and Technology, \\
Peking University, Beijing, 100871, China}

\emailAdd{pmydn@pku.edu.cn,qliphy0@pku.edu.cn}

\abstract{Triple gauge boson associated production at the LHC serves as an interesting channel to test the robustness of the Standard Model. Any deviation from its SM prediction may indicate possible existence of relevant new physics, e.g., anomalous quartic gauge boson couplings. In this paper, a Monte-Carlo feasibility study of measuring \wwa production with pure leptonic decays and probing anomalous quartic gauge-boson (e.g., \wwaa) couplings, is presented in detail for the first time, with parton shower and detector simulation effects taken into account. Our results show that at the $\sqrt{s} = 14$ TeV LHC with an integrated luminosity of 100 (30) \fbinv, one can reach a significance of 9 (5) $\sigma$ to observe the SM \wwa production, and can constrain at the 95\% CL the anomalous \wwaa coupling parameters, e.g., $a_{0,c}^W/\Lambda^2$ (see Ref.~\cite{wwaLEP:2004} for their definitions), at $1 \times 10^{-5} \text{GeV}^{\text{-2}}$, respectively.}

\date{\Date}


\begin{document}
\maketitle
\flushbottom

\section{Introduction}
\label{intr}

The Standard Model (SM) of particle physics has reached a great success below the TeV energy scale, especially after the discovery of the 125-126 GeV Higgs-like boson this year~\cite{FGianotti,JIncandela,plb:2012gu,plb:2012gk}. So far, we haven't yet observed any significant deviation from the SM. However, there are many reasons to expect new physics beyond the SM appearing at the LHC (TeV) energy scale, such as the demand for dark matter candidates and the quest to understand large hierarchy between the electroweak and Planck scales. Further confirmation of the SM or uncovering new physics beyond the SM thus become as urgent goals for both theorists and experimentalists at the LHC era.

Within the framework of the SM, gauge boson self-interactions are completely determined by the $SU(2)_{L} \otimes U(1)_{Y}$ gauge symmetry, thus direct investigation of gauge boson self-interactions provides a crucial test on the gauge structure of the SM. Moreover, since the longitudinal components of $W^\pm$ and $Z^0$ result from the spontaneously symmetry breaking, these kinds of study may also be important to explore the electroweak symmetry breaking (EWSB) mechanisms. 

For the triple gauge boson associated production which is of our interest in this paper, extra contributions other than the SM predictions can be induced by possible new physics, which can be expressed in a model independent way by high-dimensional operators which lead to anomalous triple or quartic gauge boson couplings (aTGCs or aQGCs). Compared with the gauge boson pair production channel for TGCs measurement, triple gauge boson production, although suffered from lower cross sections and more complicated final state topology, is crucial for testing QGCs, especially in the cases as discussed in e.g., Ref.~\cite{Belanger:1992qh,Bosonic:2004PRD}, that it is possible the QGCs deviate from the SM prediction while the TGCs do not,  assuming e.g., the exchange of extra heavy boson generates tree-level contributions to four gauge boson couplings while the effect on the triple gauge vertex appears only at one-loop and is consequently suppressed. 

While considerable efforts have been made on probing aTGCs, less are for aQGCs. Previous Monte-Carlo (MC) and experimental studies on aQGCs have been done at $e\gamma$ and $\gamma\gamma$ colliders~\cite{eAQGC:1993,AAQGC:1995}, linear colliders~\cite{Belanger:1992qh,zzzwwz:1996,Belanger:1999,wwaLEP:1999,zaaLEP:1999,wwaDELPHI:2003,wwaLEP:2004}, and hadron colliders~\cite{Bosonic:2004PRD,AQGC:2001,Royon:2010tw,AALHC:2003,QGCZ:1996,QGC:1995}. Direct constraints on aQGCs are available from the LEP collider searches via \wwa~\cite{wwaLEP:1999,wwaDELPHI:2003}, $\gamma\gamma \nu\nu (jj)$~\cite{wwaLEP:2004} and $Z\gamma\gamma$~\cite{zaaLEP:1999} channels, e.g., constraints at the 95\% CL on \wwaa aQGC parameters $a_{0,c}^W/\Lambda^2$  are at the order of $10^{-2} \text{GeV}^{\text{-2}}$~\cite{wwaLEP:2004}, where $a_{0,c}^W$ are dimensionless coupling constants and the $\Lambda$ stands for a new physics scale. Moreover, in Ref.~\cite{AQGC:2001}, the authors discussed that the limits on the oblique parameters S and U can be translated at 95\% CL to limits on $a_{0,c}^W/\Lambda^2$ below $1 \times 10^{-4} \text{GeV}^{\text{-2}}$ level.

With the excellent performance of the LHC and its upgrade plan to higher collision energy or luminosity in the next few years, we expect that the LHC may play a crucial role in determining the aQGCs, as shown in Refs.~\cite{Bosonic:2004PRD,AQGC:2001,Royon:2010tw} through MC simulation studies on direct \waa , VBF $\gamma\gamma$, $Z\gamma$ and $WW$ production channels, from which constraints on $a_{0,c}^W/\Lambda^2$ can be lowered to reach the order of $10^{-5}-10^{-6} \text{GeV}^{\text{-2}}$ with integrated luminosity of $30-100\fbinv$.

In this paper we focus on the MC feasibility study at the LHC of measuring \wwa production with pure leptonic decays, and probing the anomalous quartic gauge-boson coupling (\wwaa as an example), with parton shower and detector simulation effects taken into account. We begin by specifying the effective Lagrangian related to the aQGCs in Sec.~\ref{effwwaa}, and then describe the framework of our simulation studies in Sec.~\ref{Sim}, with the selection cuts demonstrated in Sec.~\ref{selec}. We present the LHC sensitivities on \wwa production with $W^+W^-$ leptonic decays and the aQGCs subsequently in Sec.~\ref{ana}. Finally we conclude in Sec.~\ref{talk}.

\section{Effective Interactions for Photonic aQGCs}
\label{effwwaa}

An effective Lagrangian can be constructed in a model independent way for the anomalous quartic couplings, assuming that new Physics beyond the SM keeps $SU(2)_{L} \otimes U(1)_{Y}$ gauge invariance. The Lagrangian can usually be written down either linearly or non-linearly~\cite{Belanger:1992qh,Belanger:1999}. The lowest order genuine aQGC operators are dim-6 for the non-linear representation and dim-8 for the linear representation~\cite{Belanger:1999,Bosonic:2004PRD}.

Although it is now more preferable to work in the linear context due to the recent discovery of Higgs-like boson, we adopt the non-linear one in order to make comparisons with the previous LEP limits. Indeed, by suitable combing various aQGC operators~\cite{Belanger:1999, Bosonic:2004PRD}, the non-linear way can lead to two same basic Lorentz structures, as the linear case, for the $WW\gamma\gamma$ aQGCs which are of our interest in this paper:

\begin{center}
\begin{equation}
{\cal W}_{0}^{\gamma} = - \frac{e^2}{8} F_{\mu \nu} F^{\mu \nu} W^{+\alpha} W^-_{\alpha} ,
\end{equation}
\end{center}

\begin{center}
\begin{equation}
{\cal W}_{c}^{\gamma} = - \frac{e^2}{16} F_{\mu \nu} F^{\mu \alpha} ( W^{+\nu} W^-_{\alpha} + W^{-\nu} W^+_{\alpha} ) ,
\end{equation}
\end{center}
 
Accordingly, the effective interactions can be expressed by the above operators as~\cite{Belanger:1999}
\begin{center}
\begin{equation}\label{operator}
{\cal L} = k_0^W {\cal W}_{0}^{\gamma} + k_c^W {\cal W}_{c}^{\gamma}.
\end{equation}
\end{center}
Here $k_{0,c}^W$ are dimensional aQGC parameters which can be compared with 
$a_{0,c}^W/\Lambda^2$ in e.g., Refs.~\cite{wwaLEP:2004,Royon:2010tw}. 

One should also note that the effective Lagrangian leads to tree-level unitarity violation at sufficiently high energy. Usually, one can regulate the rising cross section by introducing an appropriate form factor (ff). However, the choice of form factors is sort of arbitrary and can be disputable (see, e.g., \cite{Wudka:1996ah,Maestre:2011}). In this paper, we just present our results with several typical form factors for comparison, following the commonly used form factor formalism (see e.g. Refs.~\cite{Bosonic:2004PRD,Abazov:2010qn}).

\begin{eqnarray}\label{form}
a_{0,c}^W \rightarrow \frac{a_{0,c}^W}{(1 + \hat{s}/\Lambda_u^2)^n},
\end{eqnarray}
where $\hat{s}$ is the the partonic center-of-mass energy, $\Lambda_{u}$ represents the new physics scale, and in this paper we will discuss the cases of $\Lambda_u=\infty$,  $n=2$ with $\Lambda_u=1,\,2\,$TeV, and also $n=5$ with $\Lambda=2.5\,$TeV. The case of $\Lambda_u=\infty$ corresponds to no form factor attatched to the anomalous couplings as exploited at LEP~\cite{wwaLEP:1999,wwaDELPHI:2003,wwaLEP:2004,zaaLEP:1999}, which the CMS group is now also taking (for aTGC though)~\cite{Maestre:2011,Chatrchyan:2011tz}. The form factor of $n=5$ with $\Lambda=2.5\,$TeV was used in Ref.~\cite{Bosonic:2004PRD,AQGC:2001}, while the one of $n=2$ with $\Lambda=2\,$TeV in Ref.~\cite{Royon:2010tw}.

\begin{figure}{
\centering
\includegraphics[width=0.4\textwidth]{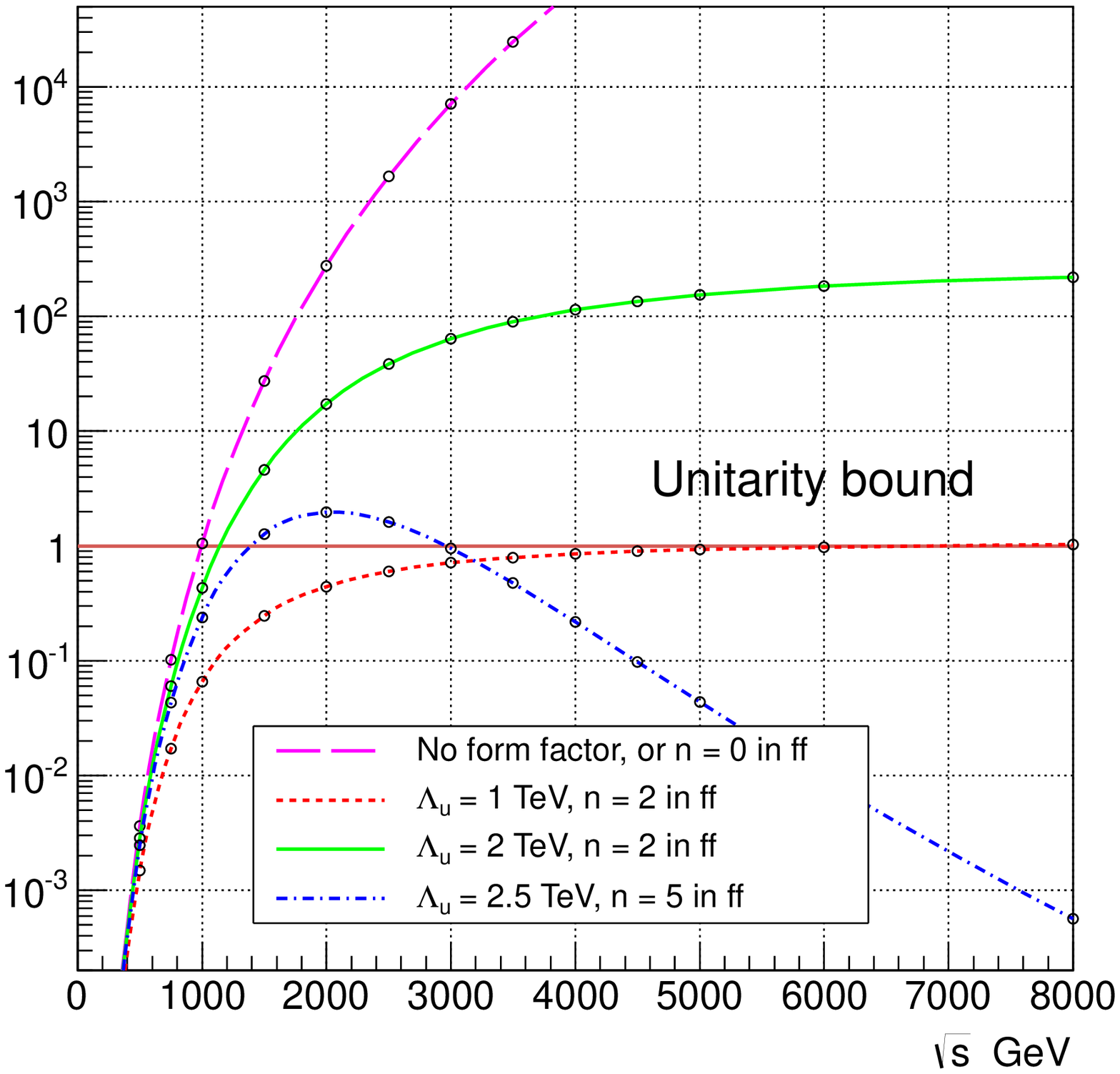}
\includegraphics[width=0.4\textwidth]{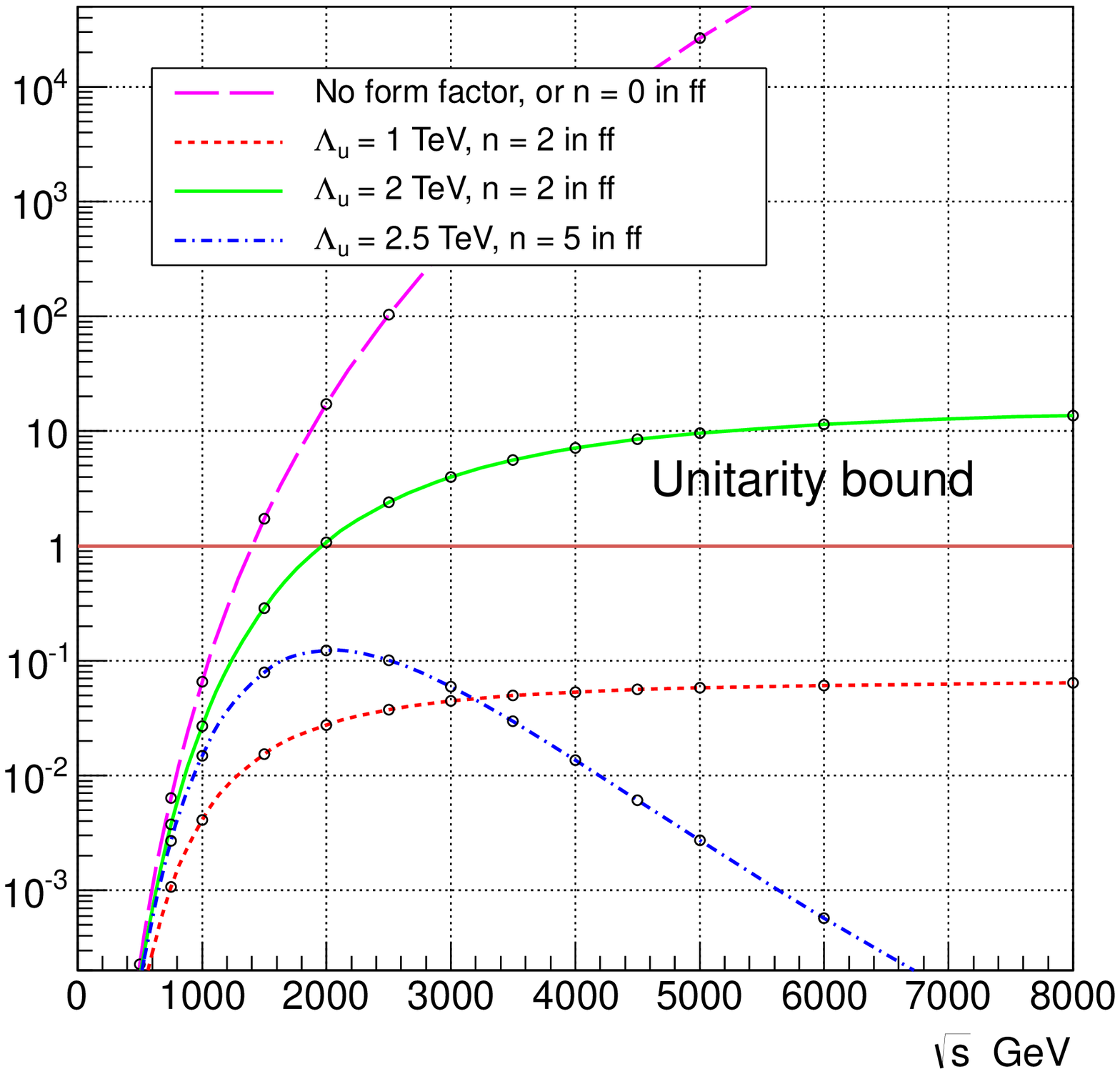}
\caption{\label{unitarity} The $J=0$ partial wave unitarity bound condition~\cite{Chapon:2009hh} for quartic anomalous coupling $a_{0}^W/\Lambda^2 = 1.5\times10^{-5}\text{GeV}^{\text{-2}}$ (left) and $a_{c}^W/\Lambda^2 = 6\times10^{-5}\text{GeV}^{\text{-2}}$ (right). The upper bound is represented by the solid lines at $y=1$. }}
\end{figure}

To further check to what extent the above mentioned choices of form factors satisfy the unitarity bounds, we make an example plot (Fig.~\ref{unitarity}) on $\hat{s}$ with $a_{0}^W/\Lambda^2 = 1.5\times10^{-5}\text{GeV}^{\text{-2}}$ and $a_{c}^W/\Lambda^2 = 6\times10^{-5}\text{GeV}^{\text{-2}}$, where the y axis value represents the left sides of the $J=0$ partial wave unitarity bound inequalities in Eqs.~(22-23) of Ref.~\cite{Chapon:2009hh}, for the anomalous interactions in Eq.~(\ref{operator}). The upper bound is represented by the solid lines at $y=1$. One can see, unitarity is violated at $\hat{s}=1 (1.4)$\,TeV if without any form factor, while at $> 4$\,TeV if with the form factor in Eq.~(\ref{form}) with $n=2$ and $\Lambda_u = 1 \text{TeV}$, for $a_{0,c}^W/\Lambda^2$, respectively. The choice of $n=5$ with $\Lambda=2.5\,$TeV also shows good behavior in general. Thus we believe our choices of the form factors (Eq.~(\ref{form})) can give meaningful and conservative estimations on the aQGC sensitivity bands, while in the meantime also provide exact comparisions with previous literatures.
 
\section{Event Simulation}
\label{Sim}

\begin{figure}{
\centering
\subfigure[With TGC]{
    \label{diagram:a}
    \includegraphics[width=0.31\textwidth]{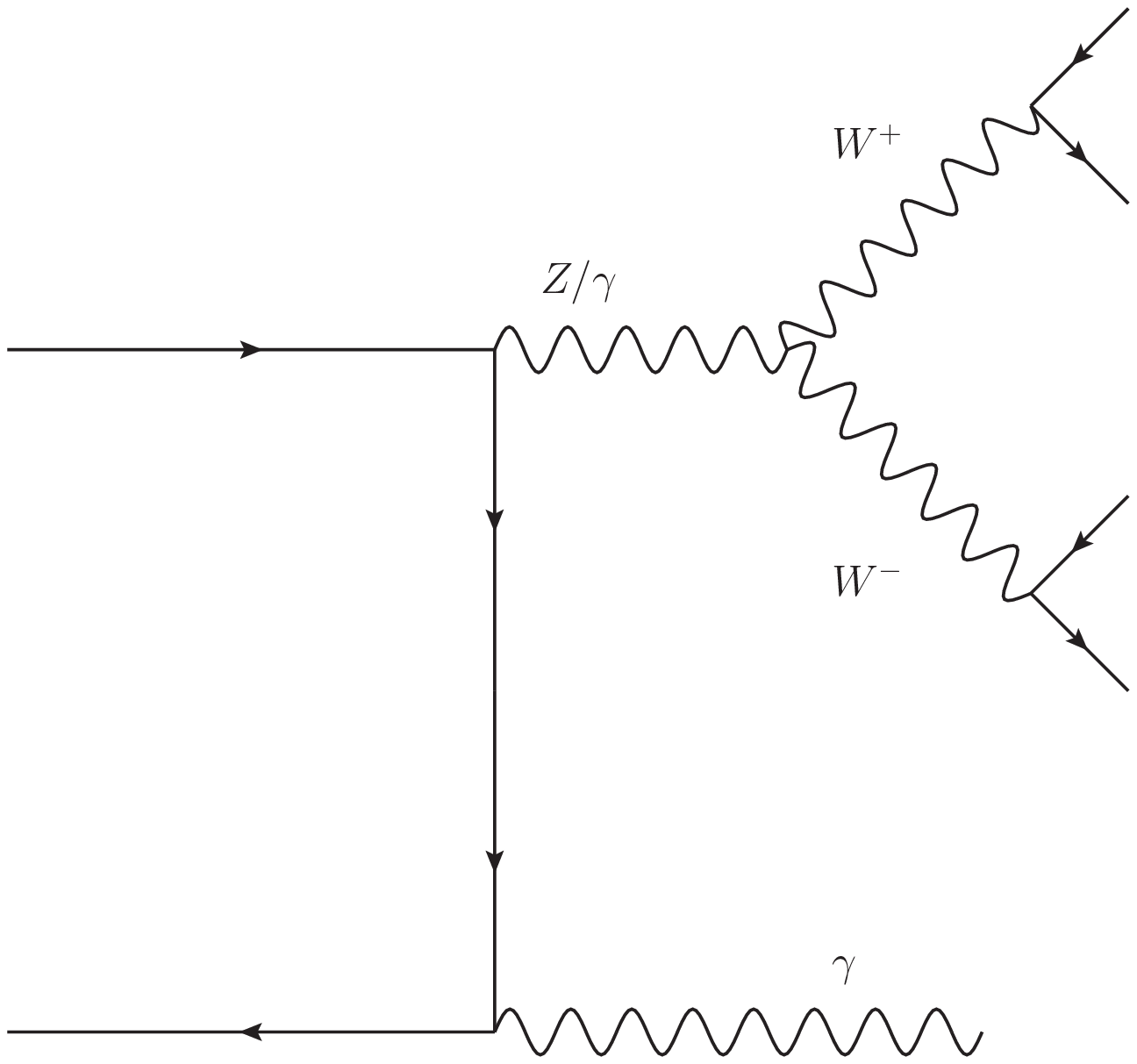}}
\subfigure[With (anomalous) QGC]{ 
    \label{diagram:b}
    \includegraphics[width=0.31\textwidth]{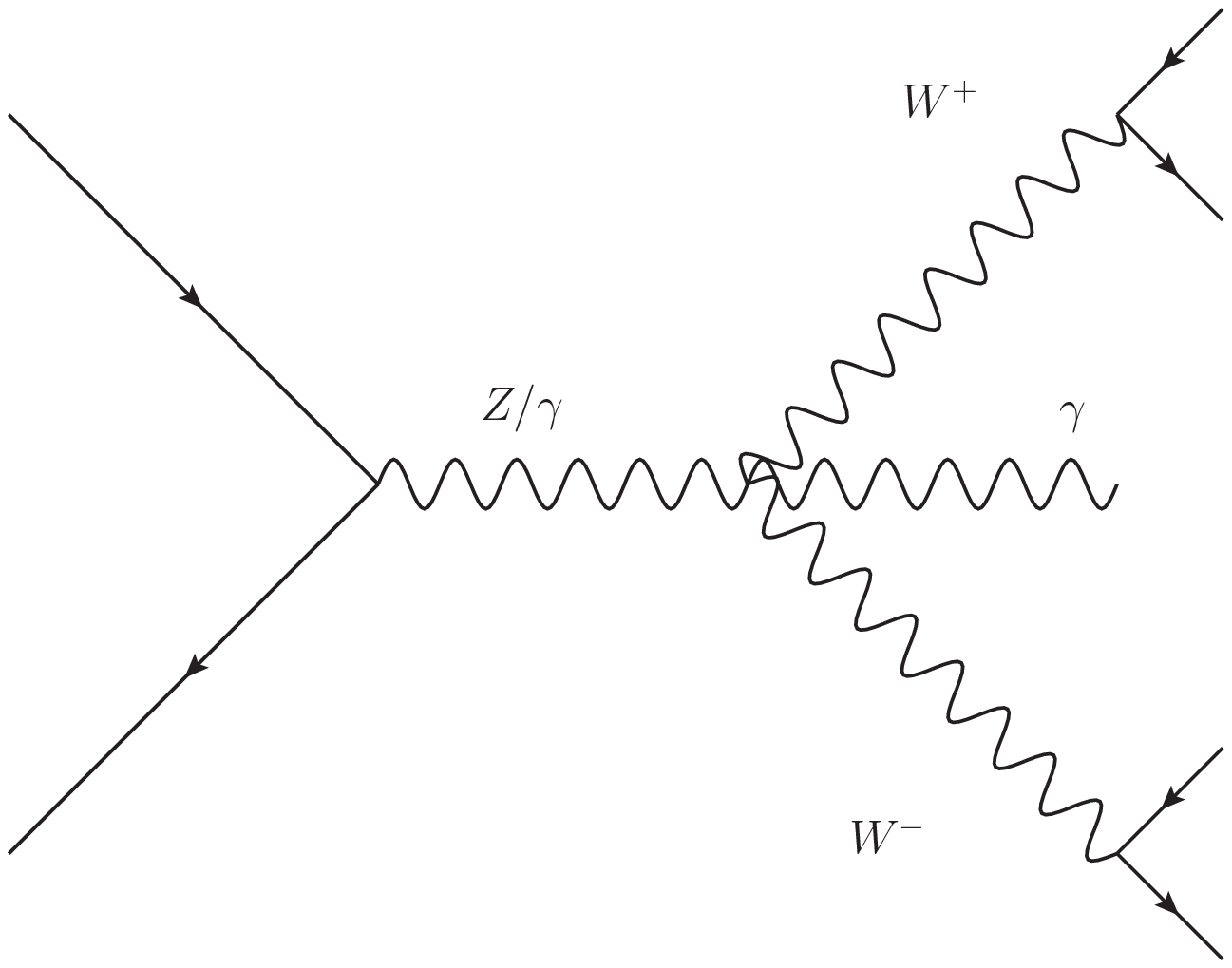}}
\subfigure[QED Radiations from $WW$]{ 
    \label{diagram:c}
    \includegraphics[width=0.31\textwidth]{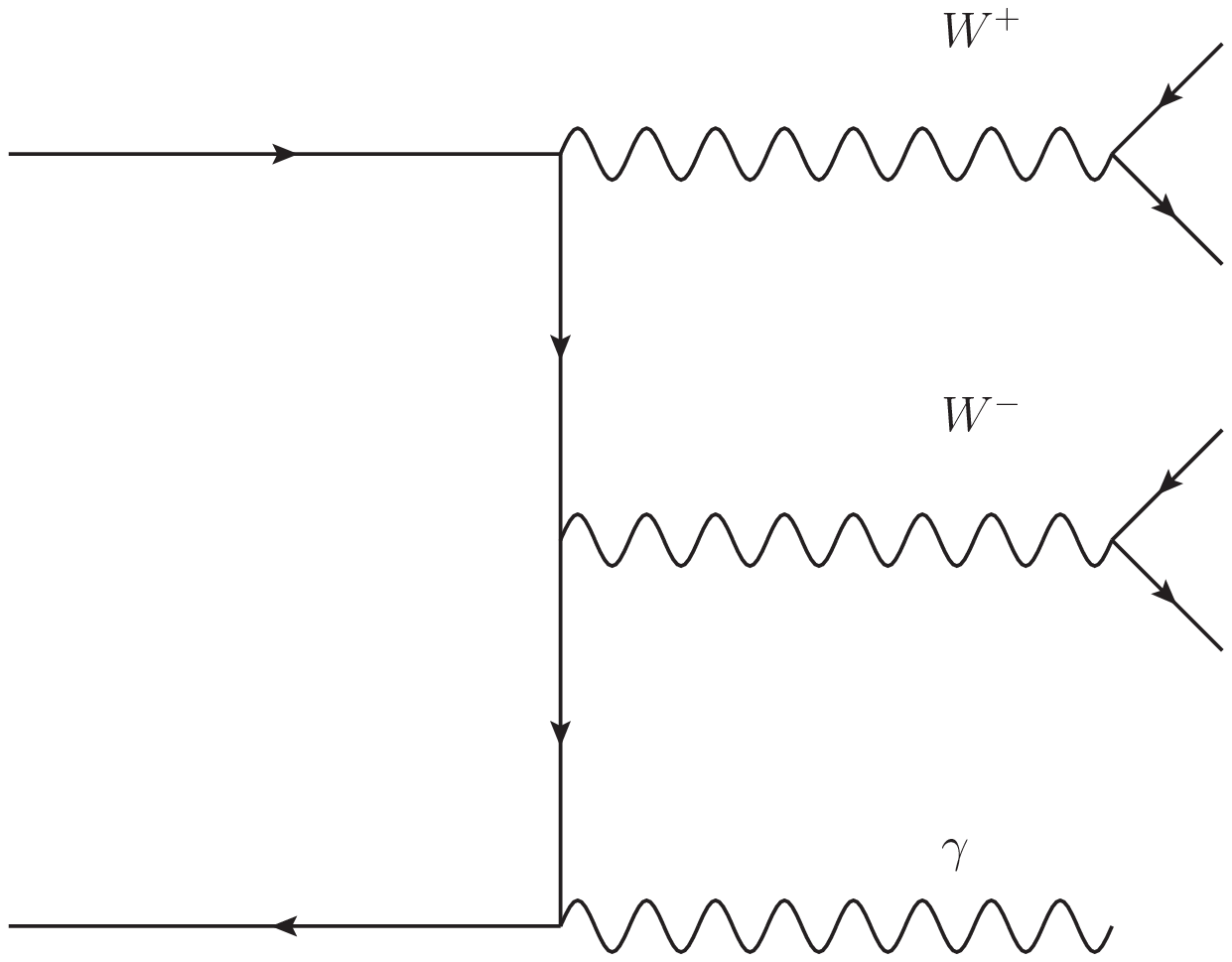}}
\caption{\label{diagrams} Example Feynman diagrams contributing to \wwa productions at the LHC}}
\end{figure}

The characteristic signal we are interested in contains two well identified lepton (electron $e$, or muon $\mu$) in association with large missing transverse energy \met . In Fig.~\ref{diagrams}, we show examples of Feynman diagrams for the \wwa productions at the LHC in the di-leptontic final state $l\nu l\nu \gamma$, with $l=e,\mu$ and $\tau$. Note $\tau$ decays into $e, \mu$ at the ratio of about 35\% and is handled with \tauola~\cite{tauola}. Five main background processes are considered: $Z \gamma$, $ZZ\gamma$, $ZW\gamma$, $tW\gamma$ and $t\bar{t}\gamma$, where $t\bar{t}\gamma$ is the dominant one. Here we don't consider backgrounds with photons from jet fragmentation, in which the photons tend to be close to jets and the contributions can be suppressed efficiently via photon isolation cuts (see e.g. Ref.~\cite{isophotons:1998}).

Attention should be paid to Fig.~\ref{diagram:c}, as it can also be seen as the initial and final state radiations (ISR and FSR) from the $WW$ production process, generated by \pythia. However, the ISR and FSR approximations in \pythia\, should break down for hard or wide scattering photon, e.g., when the transverse momentum of $\gamma$, $P_{T\,\gamma}$ is large. Note also this subset of contributions to \wwa is not related to QGCs, thus it would be interesting and important to show the overall \wwa results subtracting the contributions of the ISR/FSR approximations of Fig.~\ref{diagram:c}, which we denote as $\rm{pure\_Vs}$:
\begin{eqnarray}\label{purevs}
\rm{pure\_Vs} \equiv  WW\gamma - ISR/FSR\, WW.
\end{eqnarray}

Our work is mainly carried out under \mgme~\cite{Alwall:2007st,MadGraph:2012,Maltoni:2002qb}. 
The photonic aQGC effective Lagrangian~\ref{operator} has been implemented into \madgraph\, using the FeynRules~\cite{Christensen:2008py}-UFO~\cite{Degrande:2011ua}-ALOHA~\cite{deAquino:2011ub} framework. 
Signal and background events are then generated with \madgraph~\cite{Alwall:2007st,MadGraph:2012} and \madevent~\cite{Maltoni:2002qb}, processed subsequently by \pythia\,6~\cite{Sjostrand:2003wg} for parton showering and hadronization. Finally, the events are passed to \delphes~\cite{Ovyn:2009tx} for detector simulation, where we focus on the CMS detector at the LHC. Finally, the analysis is performed with the program package ExRootAnalysis~\cite{ExRootAnalysis} and ROOT~\cite{root}. The work flow has also been used in our previous study on semi-leptonic decayed $WW$ simulations~\cite{Liu:2012rv} .
  
\section{Event Selection}
\label{selec}
We choose the following pre-selection cuts to generate unweighted events at parton level
with \mgme\, to interface later with \pythia\, and \delphes\,, 
\begin{itemize}{
\item $P_{T\,l} \geq 15\,$GeV, $|\eta_j|<2.5$ and $R_{ll}\equiv \sqrt{\Delta\eta^2_{ll}+\Delta\phi^2_{ll}}>0.4$,  
\item $P_{T\,\gamma} \geq 10\,$GeV, and $|\eta_\gamma|<2.5$,  
\item $R_{\gamma\,l}\equiv \sqrt{\Delta\eta^2_{\gamma\,l}+\Delta\phi^2_{\gamma\,l}}>0.4$,  
}\end{itemize}
for the signals and backgrounds listed in Sec.~\ref{Sim},
where $\eta$ is the pseudo-rapidity and $\phi$ is the azimuthal angle around the beam direction. Note, however, for the backgrounds $ZZ\gamma$  and $ZW\gamma$ where one lepton is misidentified, we don't require any of the above cuts on leptons in order not to make bias.

Moreover, in the hard process generation with \mgme\, we adopt the CTEQ6L1 parton distribution functions
(PDFs)~\cite{Pumplin:2002vw} and set the renormalization and factorization scales as the transverse mass of the core process.

Tighter cuts are then imposed on the reconstructed objects in the \delphes\, settings cards,
\begin{itemize}{
\item $P_{T\,e,\mu,\gamma} \geq 20\,$GeV, and $|\eta_{e,\mu,\gamma}|<2.4$.  
\item Jets are clustered according to the anti$-k_t$ algorithm with a cone radius $\Delta R = 0.5$. 
Moreover, $P_{T,j}>P^{cut}_{T\,j}$ (25\,GeV by default) and $|\eta_{j}|<5$ are required.
}\end{itemize}

Other high level cuts are set in the analysis step as following:
\begin{itemize}{
\item (1) One and only one photon is allowed per event, and $P_{T\,\gamma}>P^{cut}_{T\,\gamma}$,
with $P^{Up}_{T\,\gamma}=40$\,GeV by default.
\item (2) Two and only two leptons with opposite sign of charge. 
\item (3) \met $> 30$\,GeV.
\item (4) $R_{\gamma\,l}$ and $R_{\gamma\,j}$ are larger than 0.5.
\item (5) To suppress top quark related backgrounds, we exclude events with b-tagged jet (with default \delphes's setting, e.g. b-tagging efficiency as $40\%$).
\item (6) To suppress backgrounds with $Z$ boson leptonic decay, we require $|m_{ll}-M_Z|>10$\,GeV.
}\end{itemize}

In \delphes, photons and charged leptons may overlap with the jet collections: \delphes\, first reconstructs photons and leptons based on MC information, and then jets which can be seeded from the already reconstructed photon or leptons. In our analysis, we clean the lepton collections from jets by requiring the \delphes's calculated ``EhadOverEem" (the energy deposition in the Hadron Calorimeter over the one in the Electromagnetic Calorimeter) smaller than 1. Moreover, we removed any jet which has $R_{j \gamma}<0.001$ as it would be indeed most like a photon.

\section{Numerical Results}
\label{ana}

\subsection{\wwa production}
\label{anawwa}

As a first step, we are interested in estimating the feasibility of observing triple gauge boson \wwa productions at the LHC, before going into aQGCs. As mentioned before, we are also interested in comparing overall \wwa results with the ISR/FSR ones from $WW$ processes (see the $WW$-column in Table~\ref{tab}).

To optimize our results, we introduce further the following 3 requirements individually, in addition to all the cuts mentioned in Sec.~\ref{selec}:

\begin{itemize}{
\item (A) Veto events with jets of which $P_{T\,j}>P^{Up}_{T\,j}\,$GeV, thus we have $P^{cut}_{T\,j}\leq P_{T\,j} \leq P^{Up}_{T\,j}$.
\item (B) Vary $P^{cut}_{T\,\gamma}$.
\item (C) Vary $P^{cut}_{T\,j}$ and require jet numbers $n_j$ to be 0 or 1. 
}\end{itemize}

We are setting the special cuts of (A) and (C), as we would like to suppress more top-quark related background while keeping high signal efficiency. Compared with signal, $t\bar{t}\gamma$ and $tW\gamma$ are different in two sides: they are QCD processes and tend to radiate more jets; the hard physics scale is higher and thus one or more jets can be harder than the jets in signal events.

\begin{figure}{
\centering
\includegraphics[width=0.31\textwidth]{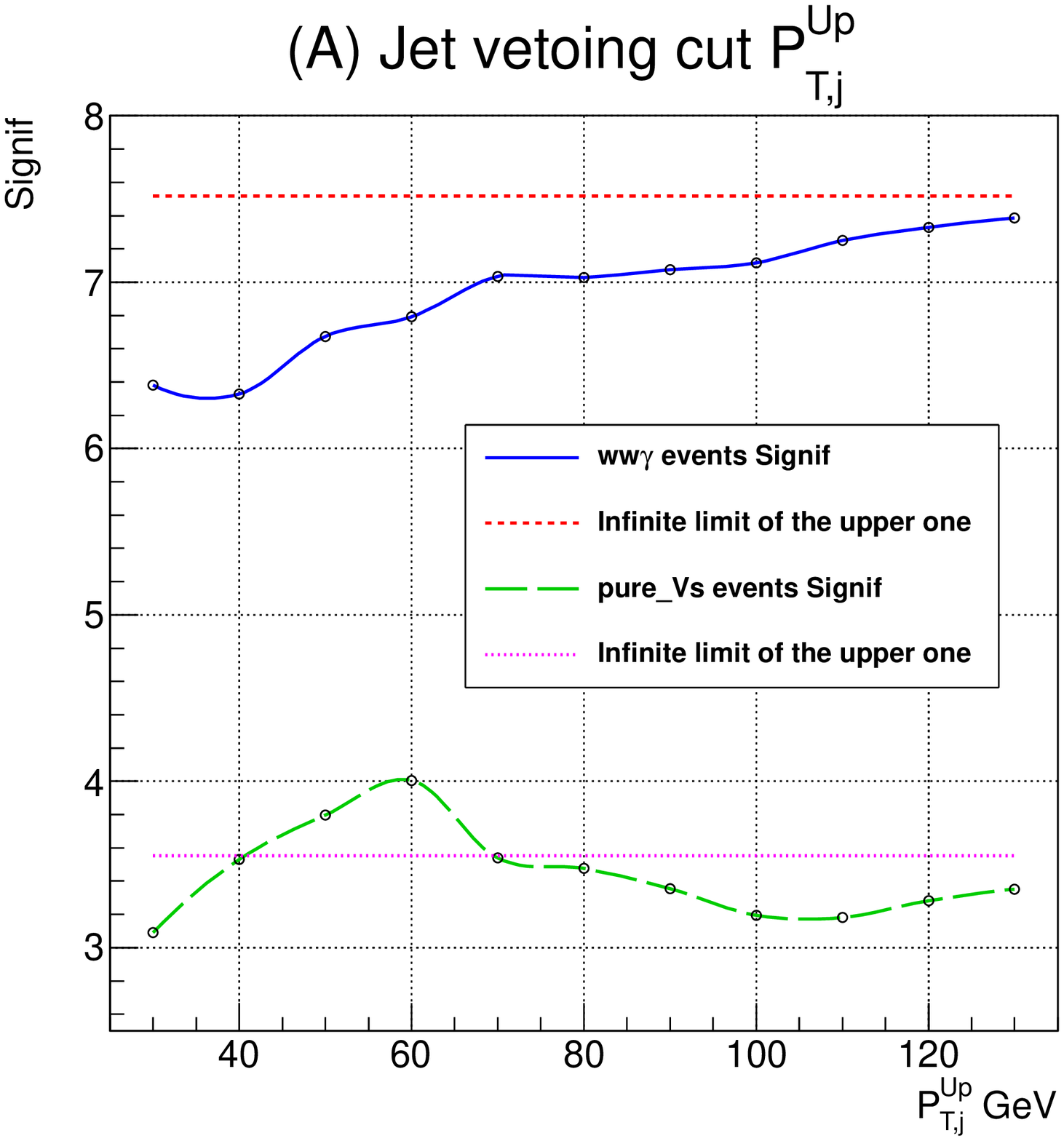}
\includegraphics[width=0.31\textwidth]{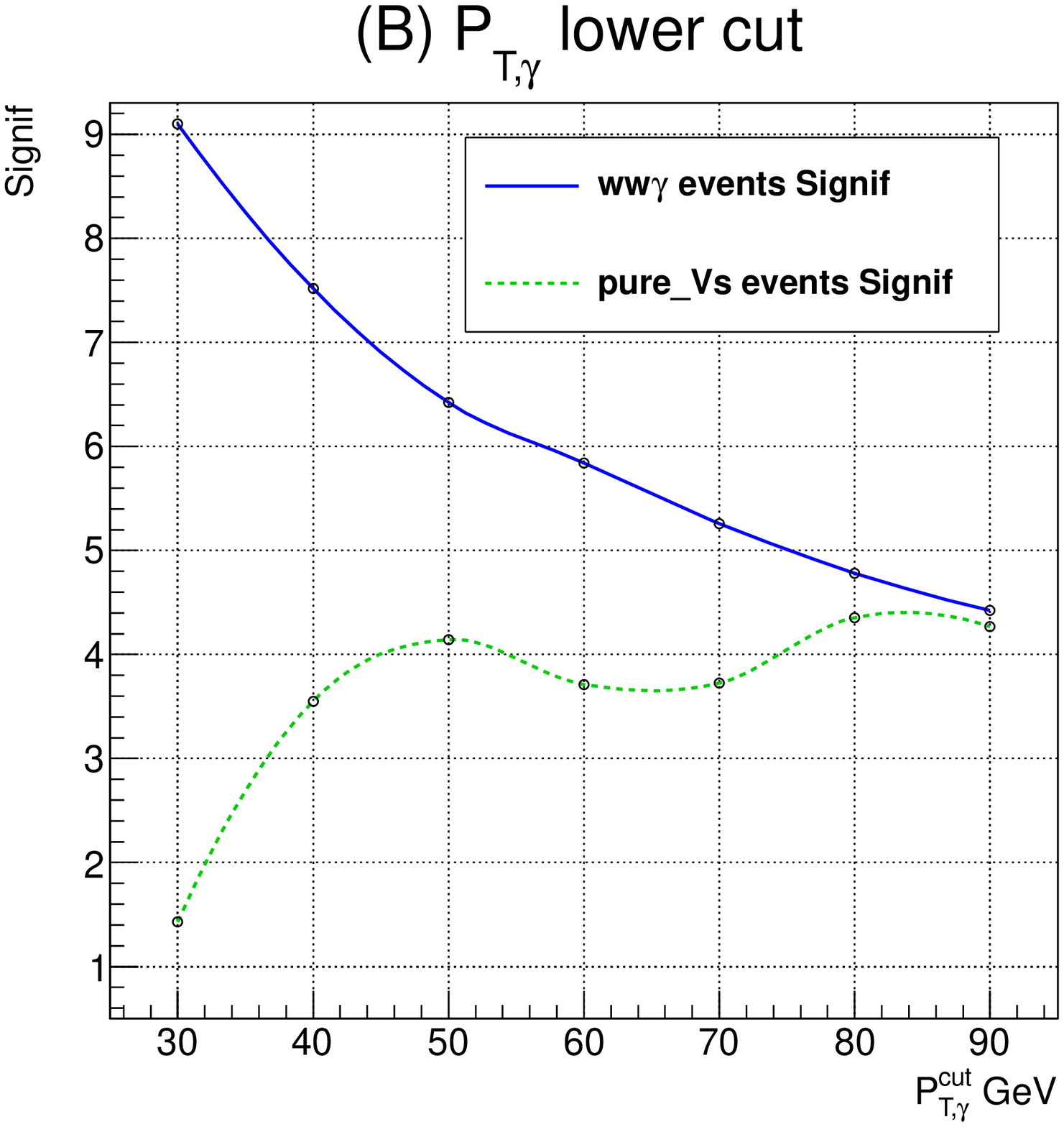}
\includegraphics[width=0.31\textwidth]{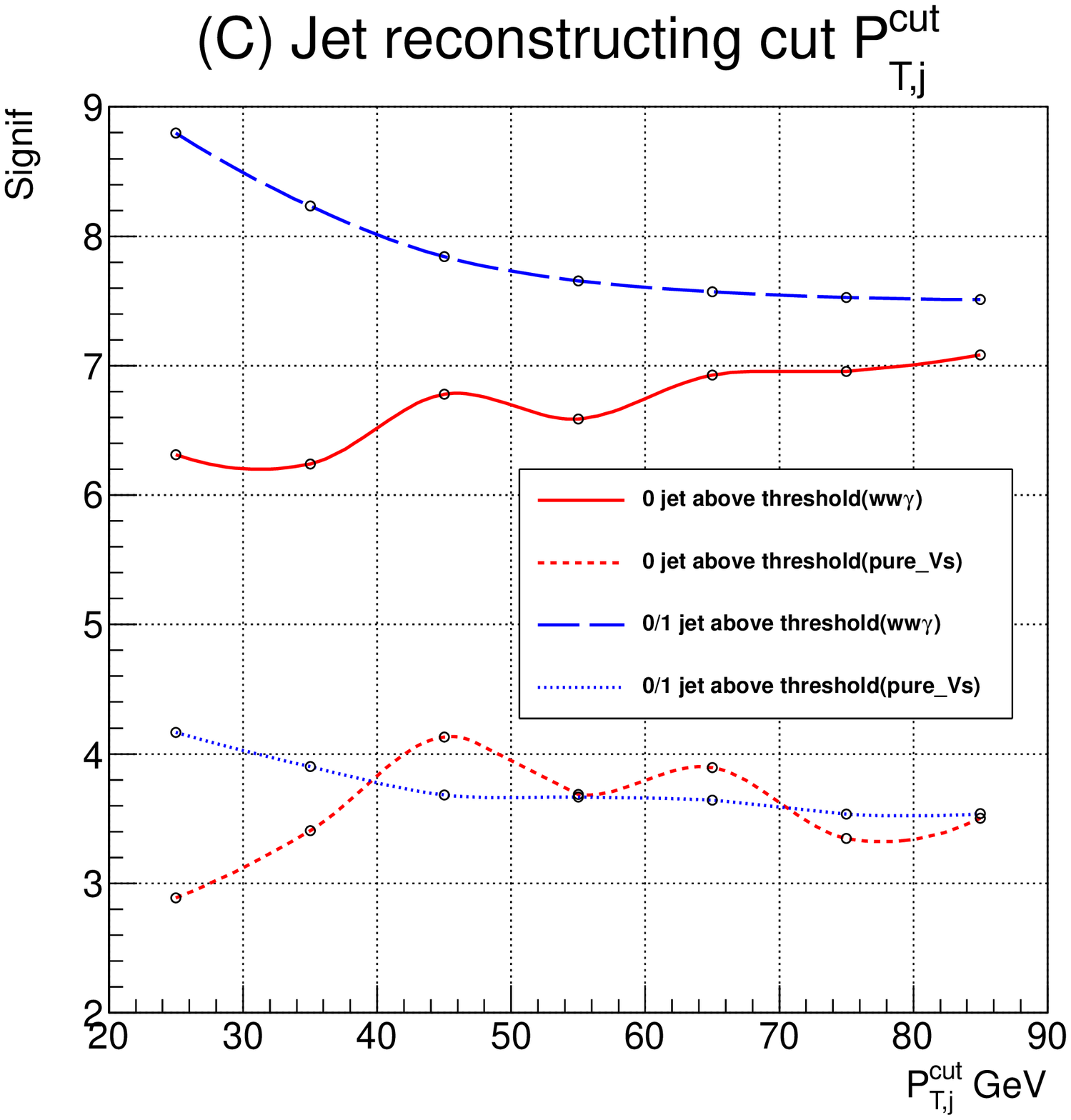}
\caption{\label{wwacuts} \wwa and $\rm{pure\_Vs}$ significances, varying Jet vetoing cut $P^{Up}_{T\,j}$, $P_{T\,\gamma}$ lower cut, and jet reconstructing cut $P^{cut}_{T\,j}$.}}
\end{figure}

We list the event numbers for the signal and backgrounds in Table~\ref{tab}, with the optimized parameters (optimized for $\rm{pure\_Vs}$ contributions) for the above 3 cases: ($A^{\ast}$) $P^{Up}_{T\,j}=60\,$GeV, ($B^{\ast}$) $P_{T\,\gamma} > 80\,$GeV, and ($C^{\ast}$) $P_{T\,j} > 25\,$GeV with $n_j\leq 1$. Related K-factors for the signal~\footnote{Note the K factor depends on $P_{T\,\gamma}$ according to Ref.~\cite{Kwwa:2009}, where it can be 1.5 for $P_{T\,\gamma} \sim 20$\,GeV. Here we take the K factor for $P_{T\,\gamma}\sim 80$\,GeV.} and backgrounds are also listed with references in Table~\ref{tab}. Correspondingly, the significances are shown in Fig.~\ref{wwacuts}, calculated with Eq.~(\ref{stat})~\cite{stat:atlas,stat:2009}.  

\begin{eqnarray}\label{stat}
Signif = \sqrt{2 ln(Q)}, \text{      } Q = (1 + N_s/N_b)^{N_{obs}} exp(-N_s), 
\end{eqnarray}

From Table~\ref{tab}), we can see that $t\bar{t}\gamma$ and $tW\gamma$ are the dominant backgrounds. It is also interesting to notice the QED ISR/FSR contributions from $WW$ get decreased a lot when $P^{cut}_{T\,\gamma}$ is set to a high value as $80$\,GeV, as the QED radiation approximation in \pythia\, breaks down for hard photons.

\begin{center}
\begin{table*}[h!]
\begin{tabular}{c||c|c||c|c|c}
\hline
\multirow{2}{*}{Processes} & Cross section & K-factor
& \multicolumn{3}{c}{ Events } \\
\cline{4-6}
& [fb] & [Ref.] & {\tiny($A^{\ast}$) $P^{Up}_{T\,j}=60$ GeV} & {\tiny($B^{\ast}$) $P^{cut}_{T\,\gamma}=80$\,GeV} & {\tiny($C^{\ast}$) $n_j=0,1$, $P_{T\,j} > 25\,$GeV}\\\hline 
\wwa              & 18.286     &  2.0~\cite{Kwwa:2009}     &     95.818      &     58.880           &     114.84            \\\hline 
I(F)SR $WW$          & 3114.1     &  1.5~\cite{mcfm}       &     35.812      &     4.6712           &     54.498            \\\hline 
$Z\gamma$         & 4107.2     &  1.5~\cite{Kza:1997}      &     61.608      &     47.232           &     57.501            \\\hline 
$ZZ\gamma$        & 45.818     &  1.3~\cite{Kwwa:2009}     &     0.2779      &     0.1985           &     0.2780            \\\hline 
$W^{\pm}Z\gamma$  & 1.3698     &  1.5~\cite{Kwza:2010}     &     0.8903      &     0.5739           &     1.0068            \\\hline 
$t\bar{t}\gamma$  & 170.22     &  1.9~\cite{Kttbara:2011}  &     88.830      &     73.738           &     60.801            \\\hline 
$tW^{\pm}\gamma$  & 26.858     &  1.0~\cite{mcfm}          &     17.905      &     11.442           &     16.527            \\\hline 
\end{tabular}
\caption{\label{tab} Cut flow at the LHC with $\sqrt{s}=14$ TeV and integrated luminosity of $100\fbinv$.}
\end{table*}
\end{center}

More details can also be checked in Fig.~\ref{wwacuts}, showing the significance dependences on (A) jet vetoing cut $P^{Up}_{T\,j}$, (B) $P^{cut}_{T\,\gamma}$, and (C) $P^{cut}_{T\,j}$. Note we also give the $\rm{pure\_Vs}$-curves to show the results after subtracting ISR/FSR contributions from $WW$ processes, as mentioned above.

(A). $t\bar{t}\gamma$ and $tW\gamma$ backgrounds tend to have harder jet than the signal. Increasing the Jet vetoing cut $P^{Up}_{T\,j}$ from low value as 30 GeV, at first can veto more top background while less signal events, and thus enhance the significance for both \wwa and $\rm{pure\_Vs}$ cases. However, the gain gets small when $P^{Up}_{T\,j}$ is as large as 60 to 100 GeV (as top decayed jets are now soft compared with $P^{Up}_{T\,j}$), for $\rm{pure\_Vs}$ and \wwa, respectively.  

(B). Increasing $P_{T\,\gamma}$ lower cut decreases the overall significance of \wwa, as the events get decreased for both signal and backgrounds. However, the $\rm{pure\_Vs}$-significance increased, as the percentage of $\rm{pure\_Vs}$ over \wwa is enhanced.

(C). Increasing jet reconstructing cut $P^{cut}_{T\,j}$, increases the 0-jet contributions while decreases the 1-jet ones, as expected. The overall 0+1 jet significances are decreased, as less $t\bar{t}\gamma$ and $tW\gamma$ events (which have more QCD radiations than the signal) are discarded.

Above all, a high significance as 7 to 9$\sigma$ can be achieved to observe $WW\gamma$ productions at the 14 TeV LHC, depending on the cuts. e.g., 7.4$\sigma$, 9.1$\sigma$ and 8.8$\sigma$ for cases ($A$), ($B$) and ($C$), respectively. Note a large portion of $WW\gamma$ events can come from the QED ISR/FSR $WW$ which is not related to QGCs, as shown by the $\rm{pure\_Vs}$-curves in Fig.~\ref{wwacuts}, however, sticking to large $P_{T\,\gamma}$ lower cut ($\sim 90$\,GeV), one can still get a high significance about 4.5 sigma and almost totally from $\rm{pure\_Vs}$ contributions.

\subsection{Anomalous \wwaa Couplings}
\label{anowwaa}

\begin{figure}{
\centering
\includegraphics[width=0.4\textwidth]{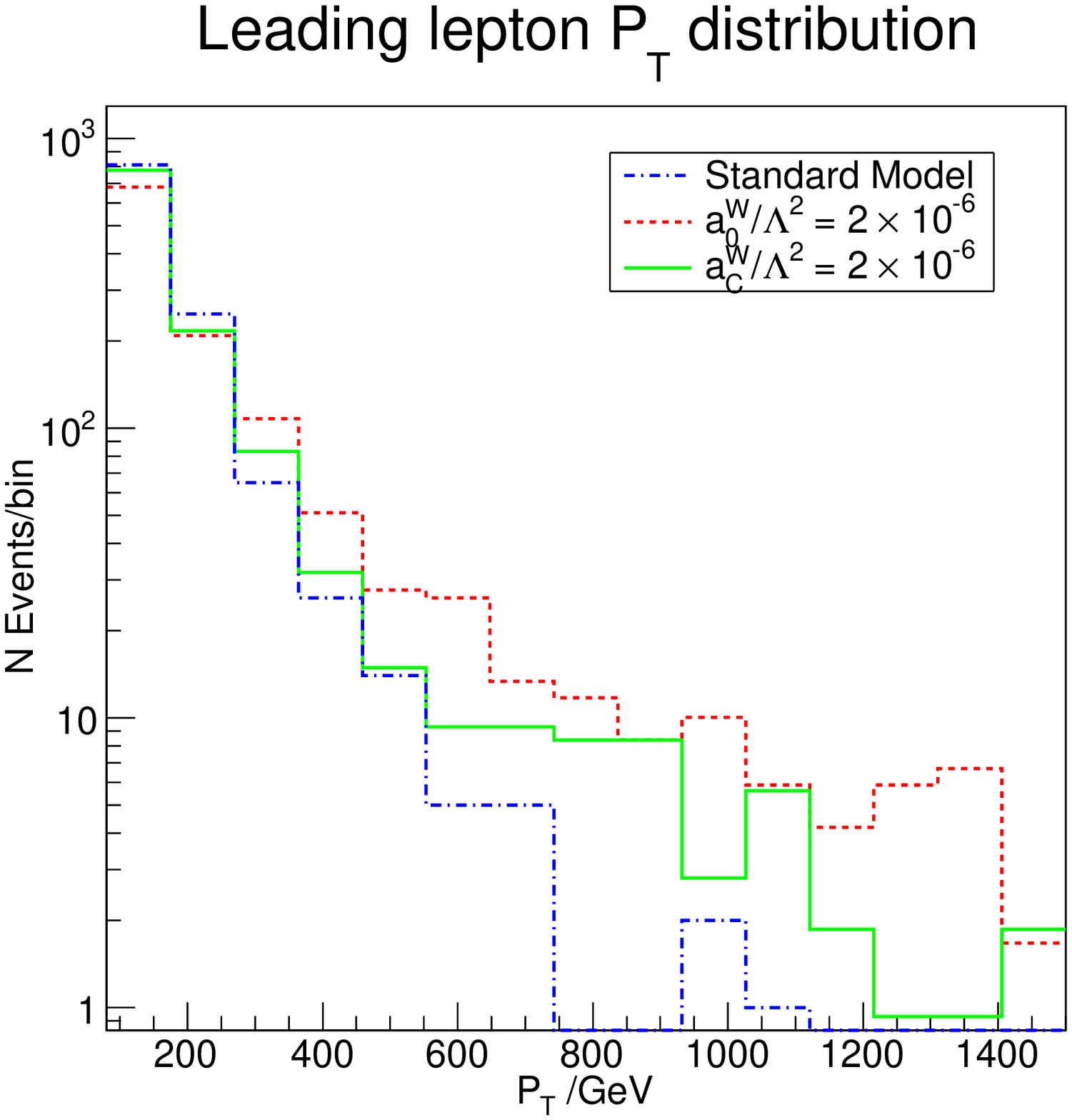}
\includegraphics[width=0.4\textwidth]{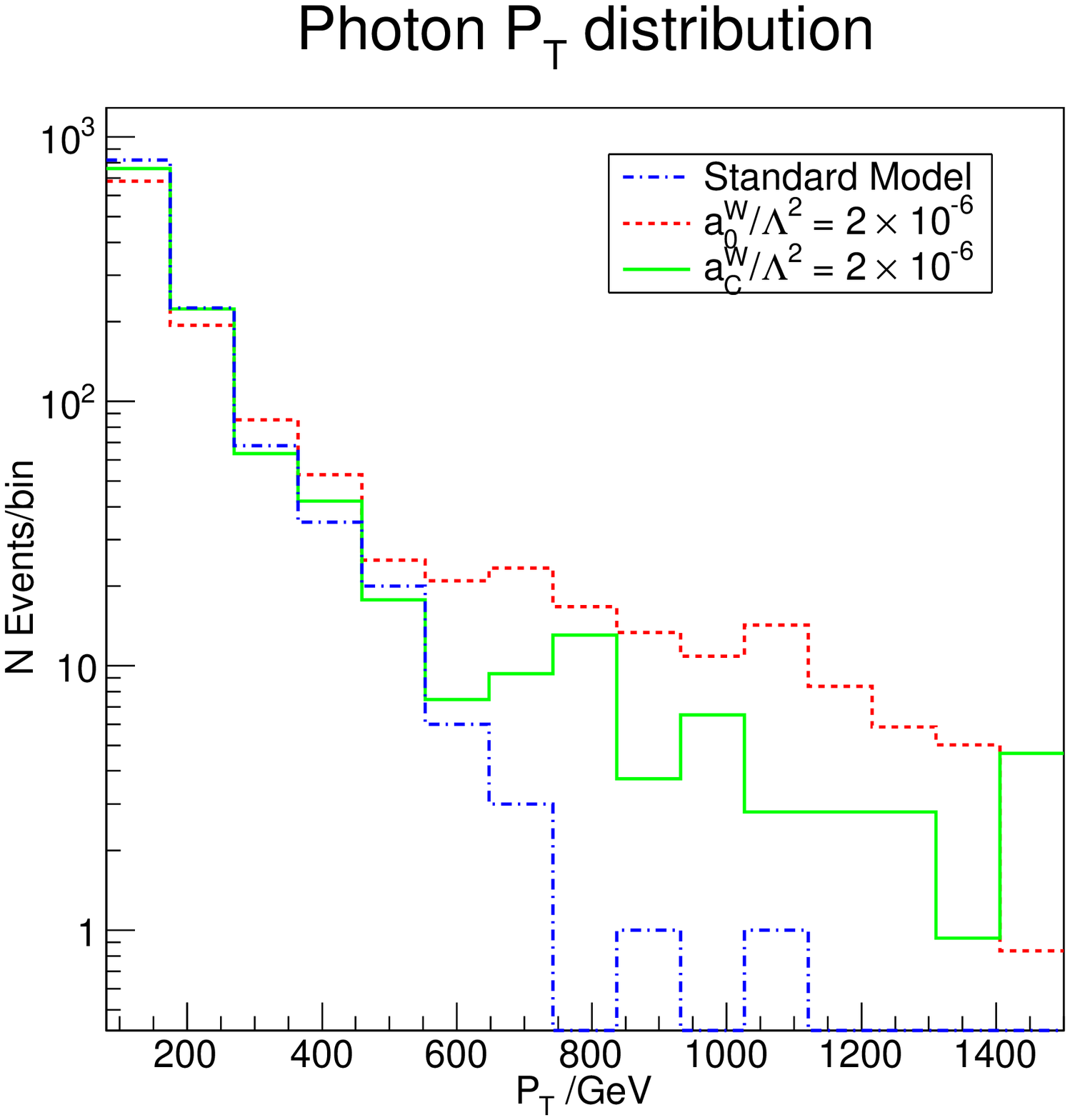}
\caption{\label{ancut} Comparing the differential distributions for $WW\gamma$ productions at the LHC in leading lepton $P_T$ and Photon $P_T$, with the SM \wwaa QGC, or aQGCs $a_{0,c}^W/\Lambda^2 = 2\times10^{-6}\text{GeV}^{\text{-2}}$. No Form factor is applied here.}}
\end{figure}

The \wwa signal process can be sensitive to aQGCs \wwaa and \wwza. In this paper we are only considering \wwaa aQGCs for simplicity. The cross sections (via \mgme\, after the pre-selection cuts mentioned in Sec.~\ref{selec}) of \wwa productions at the LHC can grow quickly with the increase of the absolute values of aQGCs, as demonstrated in Fig.~\ref{anXS}. Moreover, As shown in Fig.~\ref{ancut}, the aQGCs lead to excesses on the hard tails in various kinematic region. One thus can refine the cuts in Sec.~\ref{selec} to enhance the sensitivity to QGCs as following, e.g. :

\begin{figure}{
\centering
\includegraphics[width=0.4\textwidth]{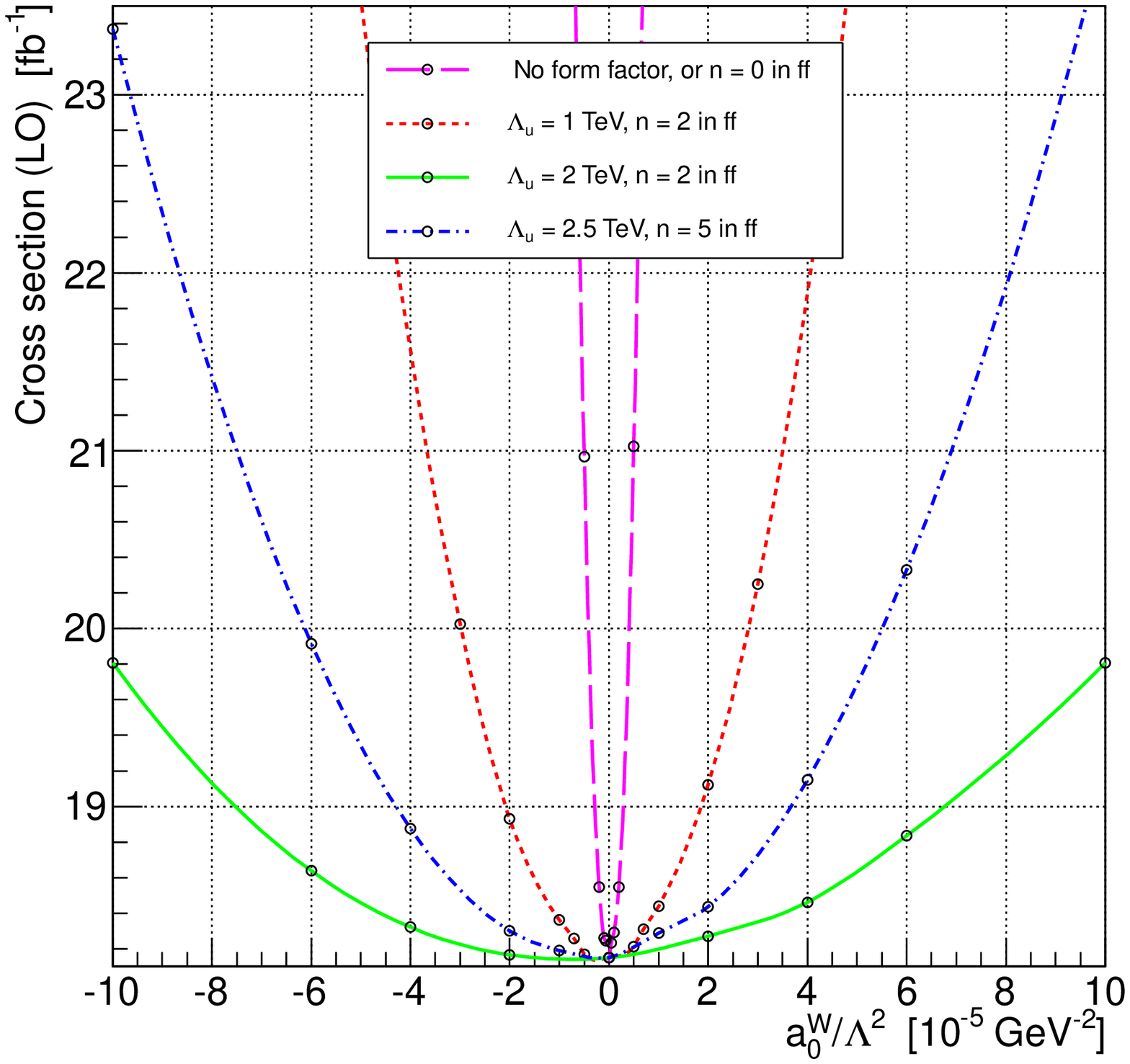}
\includegraphics[width=0.4\textwidth]{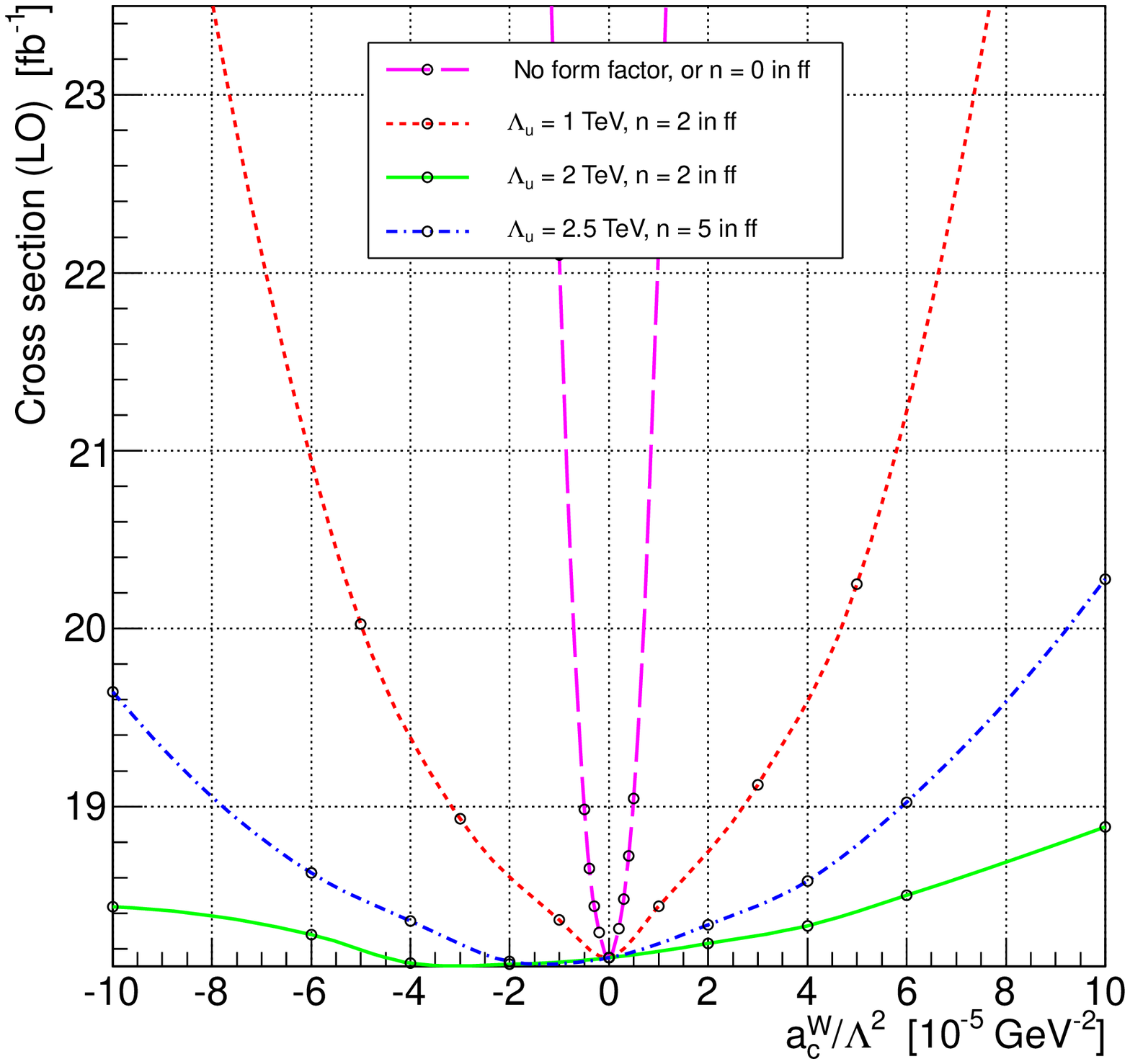}
\caption{\label{anXS} Cross section dependences on \wwaa anomalous couplings $a_{0,c}^W/\Lambda^2$ at the $\sqrt{s} = 14$ TeV LHC, with the form factors introduced in Eq.~(\ref{form}).}}
\end{figure}

\begin{itemize}{
\item (a) $P_{T\,\gamma}>250$\,GeV.
\item (b) The leading lepton $P_T>200$\,GeV. 
\item (c) $|m_{ll}-M_Z|>5$\,GeV.
}\end{itemize}

After all these selection cuts, the significances are calculated and displayed as following as functions of the QGCs $k_{0,c}^W$, at the 14 TeV LHC, with an integrated luminosity of 30(100)[200] \fbinv, respectively. As mentioned previously in Sec.~\ref{effwwaa}, the results are presented with 4 different kinds of form factors:

\begin{itemize}{
\item (a) No form factor, $\Lambda_u = \infty $ or $n = 0$,
\begin{center}
\begin{eqnarray}
-0.089{(-0.073)}[-0.068] \times 10^{-5} \text{GeV}^{\text{-2}} < a_0^W/\Lambda^2 < 0.179{(0.136)}[0.116] \times 10^{-5} \text{GeV}^{\text{-2}}, \quad\quad \label{anopkog1}\\
-0.310{(-0.235)}[-0.215] \times 10^{-5} \text{GeV}^{\text{-2}} < a_c^W/\Lambda^2 < 0.310{(0.235)}[0.205] \times 10^{-5} \text{GeV}^{\text{-2}}. \quad\quad \label{anopkcg1}
\end{eqnarray}
\end{center}

\item (b) ff with $\Lambda_u = 2$ TeV, $n = 2$,
\begin{center}
\begin{eqnarray}
-1.40(-0.82)[-0.66] \times 10^{-5} \text{GeV}^{\text{-2}} < a_0^W/\Lambda^2 < 1.42(1.04)[0.66] \times 10^{-5} \text{GeV}^{\text{-2}}, \quad\quad \label{anopkog2}\\
-2.60(-1.86)[-1.62] \times 10^{-5} \text{GeV}^{\text{-2}} < a_c^W/\Lambda^2 < 2.26(1.60)[1.24] \times 10^{-5} \text{GeV}^{\text{-2}}. \quad\quad \label{anopkcg2}
\end{eqnarray}
\end{center}

\item (c)  ff with $\Lambda_u = 1$ TeV, $n = 2$,
\begin{center}
\begin{eqnarray}
-6.8(-4.9)[-4.2] \times 10^{-5} \text{GeV}^{\text{-2}} < a_0^W/\Lambda^2 < 6.5(5.1)[4.7] \times 10^{-5} \text{GeV}^{\text{-2}}, \label{anopkog3} \\
-10.8(-9.3)[-8.3] \times 10^{-5} \text{GeV}^{\text{-2}} < a_c^W/\Lambda^2 < 10.6(7.2)[5.3] \times 10^{-5} \text{GeV}^{\text{-2}}. \label{anopkcg3}
\end{eqnarray}
\end{center}

\item (d)  ff with $\Lambda_u = 2.5$ TeV, $n = 5$,
\begin{center}
\begin{eqnarray} 
-3.45(-2.56)[-2.15] \times 10^{-5} \text{GeV}^{\text{-2}} < a_0^W/\Lambda^2 < 3.55(1.99)[1.64] \times 10^{-5} \text{GeV}^{\text{-2}}, \quad\quad \label{anopkog4} \\
-6.8(-5.05)[-2.95] \times 10^{-5} \text{GeV}^{\text{-2}} < a_c^W/\Lambda^2 < 5.92(3.5)[2.85] \times 10^{-5} \text{GeV}^{\text{-2}}. \quad\quad \label{anopkcg4}
\end{eqnarray}
\end{center}

}\end{itemize}
 
One can also compare our results Eqs.~(\ref{anopkog1})-(\ref{anopkcg1}), Eqs.~(\ref{anopkog2})-(\ref{anopkcg2}), and Eqs.~(\ref{anopkog4})-(\ref{anopkcg4}), with the existing constraints given by the OPAL Collaboration~\cite{wwaLEP:2004} and the previous MC expectation limits based on \waa~\cite{AQGC:2001}, VBF photon exchange~\cite{Royon:2010tw}, and W Boson Fusion (WBF) processes~\cite{Bosonic:2004PRD}~\footnote{Note Ref.~\cite{Bosonic:2004PRD} used different parametrization scheme and we have tranlated limits listed therein for our case.}, respectively, as shown in Table~\ref{comptab}. At the 14 TeV LHC, we can set more stringent limit than the OPAL one, down to $1 \times 10^{-5} \text{GeV}^{\text{-2}}$ level with only 30\fbinv LHC data. Our results also show that \wwa channel can compete with or work better than \waa~\cite{AQGC:2001} and VBF W boson fusion~\cite{Bosonic:2004PRD}. Although our \wwa channel seems to set more loose limits on QGCs than the VBF photon exchange process~\cite{Royon:2010tw}, however, it has simpler event topology and may be less contaminated by the QCD and VBF systematics.

\begin{center}
\begin{table*}[h!]
\begin{tabular}{c||c|c|c|c}
\hline
Couplings & OPAL Limit & \waa & W boson fusion & Photon exchange \\
\cline{3-5}
 [$\text{GeV}^{\text{-2}}$] &   & 100 \fbinv & 100 \fbinv & 30{(200)}\fbinv  $\:$ \\\hline 

$a_0^W/\Lambda^2$ & [-0.020, 0.020] & [-7.6, 7.6] $\times 10^{-5}$ & 1.4 $\times 10^{-5}$ & 0.26 (0.14) $\times 10^{-5}$       \\\hline 
$a_c^W/\Lambda^2$ & [-0.052, 0.037] & [-11, 10] $\times 10^{-5}$ & 5.3 $\times 10^{-5}$ & 0.94 (0.52) $\times 10^{-5}$       \\\hline 

\end{tabular}
\caption{\label{comptab} 95\% C.L. limits for existing LEP2 results~\cite{wwaLEP:2004}, the \waa~\cite{AQGC:2001}, VBF W boson fusion~\cite{Bosonic:2004PRD} and photon exchange~\cite{Royon:2010tw} processes, which can be compared with our results with the same form factors, Eqs.~(\ref{anopkog1})-(\ref{anopkcg1}), Eqs.~(\ref{anopkog2})-(\ref{anopkcg2}), and Eqs.~(\ref{anopkog4})-(\ref{anopkcg4}), respectively.}
\end{table*}
\end{center}

\section{Discussion}
\label{talk}

In the past, due to the insufficiency of the available center-of-mass energy, measurement on triple boson production has never been performed at hadron colliders. With the excellent behavior and possible upgration in the near future of the powerful LHC, this kind of measurement could become possible in the future. 

In summary, our study shows that at the 14 TeV LHC with an integrated luminosity of 100 (30) \fbinv, one can reach a significance of 9.1 (5.0) $\sigma$ to observe the SM \wwa production, and can constrain at the 95\% CL the anomalous \wwaa coupling parameters, e.g., $a_{0,c}^W/\Lambda^2$ (see Ref.~\cite{wwaLEP:2004} for their definitions), at $1 \times 10^{-5} \text{GeV}^{\text{-2}}$, respectively. The expected limits are far beyond the existing LEP results, and can be comparable with the ones from \waa~\cite{AQGC:2001} and VBF W boson fusion~\cite{Bosonic:2004PRD}, although less tighter than the ones from  VBF photon exchange processes~\cite{Royon:2010tw}, suffer less the QCD and VBF systematics due to cleaner and simpler events topology.

\acknowledgments
Daneng Yang would like to thank Wei Shan for helpful discussions and pointing out good references, and Jiangbo Wei for helping us solving some editing problems. This work is supported in part by the National Natural Science Foundation of China, under Grants No. 10721063, No. 10975004, No. 10635030 and No. 11205008.


\appendix



\begin{thebibliography}{00}
\bibliographystyle{apsrev}
\bibliography{references}


\bibitem{FGianotti} 
F.~Gianotti, CERN Seminar, ”Update on the Standard Model Higgs searches in AT-
LAS”, July, 4 2012. ATLAS-CONF-2012-093

\bibitem{JIncandela} 
J.~Incandela, CERN Seminar, ”Update on the Standard Model Higgs searches in CMS”,
July, 4 2012.

\bibitem{plb:2012gu} 
  S.~Chatrchyan {\it et al.}  [CMS Collaboration],
  Phys.\ Lett.\ B {\bf 716}, 30 (2012)
  [arXiv:1207.7235 [hep-ex]].

\bibitem{plb:2012gk} 
  G.~Aad {\it et al.}  [ATLAS Collaboration],
  Phys.\ Lett.\ B {\bf 716}, 1 (2012)
  [arXiv:1207.7214 [hep-ex]]. 

\bibitem{Belanger:1992qh} 
  G.~Belanger and F.~Boudjema,
  Phys.\ Lett.\ B {\bf 288}, 201 (1992).

\bibitem{Bosonic:2004PRD}
  O.~J.~P.~Eboli, M.~C.~Gonzalez-Garcia and S.~M.~Lietti,
  Phys.\ Rev.\ D {\bf 69}, 095005 (2004)
  [hep-ph/0310141].

\bibitem{Chapon:2009hh} 
  E.~Chapon, C.~Royon and O.~Kepka,
  Phys.\ Rev.\ D {\bf 81}, 074003 (2010)
  [arXiv:0912.5161 [hep-ph]].




\bibitem{eAQGC:1993}
  O.~J.~P.~Eboli, M.~C.~Gonzalez-Garcia and S.~F.~Novaes,
  Nucl.\ Phys.\ B {\bf 411}, 381 (1994)
  [hep-ph/9306306].

\bibitem{AAQGC:1995}
  O.~J.~P.~Eboli, M.~B.~Magro, P.~G.~Mercadante and S.~F.~Novaes,
  Phys.\ Rev.\ D {\bf 52}, 15 (1995)
  [hep-ph/9503432].

\bibitem{zzzwwz:1996} 
  S.~Dawson, A.~Likhoded, G.~Valencia and O.~Yushchenko,
  eConf C {\bf 960625}, NEW147 (1996)
  [hep-ph/9610299].

\bibitem{Belanger:1999}
  G.~Belanger, F.~Boudjema, Y.~Kurihara, D.~Perret-Gallix and A.~Semenov,
  Eur.\ Phys.\ J.\ C {\bf 13}, 283 (2000)
  [hep-ph/9908254].

\bibitem{wwaLEP:1999} 
  G.~Abbiendi {\it et al.}  [OPAL Collaboration],
  Phys.\ Lett.\ B {\bf 471}, 293 (1999)
  [hep-ex/9910069].

\bibitem{zaaLEP:1999} 
  M.~Acciarri {\it et al.}  [L3 Collaboration],
  Phys.\ Lett.\ B {\bf 478}, 39 (2000)
  [hep-ex/0002037].

\bibitem{wwaDELPHI:2003} 
  J.~Abdallah {\it et al.}  [DELPHI Collaboration],
  Eur.\ Phys.\ J.\ C {\bf 31}, 139 (2003)
  [hep-ex/0311004].

\bibitem{wwaLEP:2004} 
  G.~Abbiendi {\it et al.}  [OPAL Collaboration],
  Phys.\ Rev.\ D {\bf 70}, 032005 (2004)
  [hep-ex/0402021].

\bibitem{AQGC:2001}
  O.~J.~P.~Eboli, M.~C.~Gonzalez-Garcia, S.~M.~Lietti and S.~F.~Novaes,
  Phys.\ Rev.\ D {\bf 63}, 075008 (2001)
  [hep-ph/0009262].

\bibitem{Royon:2010tw} 
  C.~Royon, E.~Chapon and O.~Kepka,
  PoS DIS {\bf 2010}, 089 (2010)
  [AIP Conf.\ Proc.\  {\bf 1350}, 140 (2011)]
  [arXiv:1008.0258 [hep-ph]].

\bibitem{AALHC:2003}
  T.~Pierzchala and K.~Piotrzkowski,
  Nucl.\ Phys.\ Proc.\ Suppl.\  {\bf 179-180}, 257 (2008)
  [arXiv:0807.1121 [hep-ph]].

\bibitem{QGCZ:1996}
  A.~Brunstein, O.~J.~P.~Eboli and M.~C.~Gonzalez-Garcia,
  Phys.\ Lett.\ B {\bf 375}, 233 (1996)
  [hep-ph/9602264].

\bibitem{QGC:1995}
  S.~Godfrey,
  In *Los Angeles 1995, Vector boson self-interactions* 209-223
  [hep-ph/9505252].

\bibitem{Wudka:1996ah} 
  J.~Wudka,
  eConf C {\bf 960625}, NEW176 (1996)
  [hep-ph/9606478].


\bibitem{Maestre:2011}
Juan Alcaraz Maestre, Experimental limits on anomalous TGC couplings and
future plans, Implications of LHC results for TeV-scale physics, CERN, Aug. 29 -Sep. 2, 2011.

\bibitem{Abazov:2010qn} 
  V.~M.~Abazov {\it et al.}  [D0 Collaboration],
  Phys.\ Lett.\ B {\bf 695}, 67 (2011)
  [arXiv:1006.0761 [hep-ex]].

\bibitem{Chatrchyan:2011tz} 
  S.~Chatrchyan {\it et al.}  [CMS Collaboration],
  Phys.\ Lett.\ B {\bf 699}, 25 (2011)
  [arXiv:1102.5429 [hep-ex]].
 
\bibitem{isophotons:1998}
  S.~Frixione,
  Phys.\ Lett.\ B {\bf 429}, 369 (1998)
  [hep-ph/9801442].

\bibitem{tauola} 
Jadach, Stanislaw et al. Comput.\ Phys.\ Commun.\ 64, 275 (1990). CERN-TH-5856-90. 

\bibitem{Alwall:2007st}
  J.~Alwall {\it et al.},
  JHEP {\bf 0709}, 028 (2007)
  [arXiv:0706.2334 [hep-ph]].

\bibitem{MadGraph:2012}
  J.~Alwall, M.~Herquet, F.~Maltoni, O.~Mattelaer and T.~Stelzer,
  JHEP {\bf 1106} (2011) 128
  [arXiv:1106.0522 [hep-ph]].

\bibitem{Christensen:2008py} 
  N.~D.~Christensen and C.~Duhr,
  Comput.\ Phys.\ Commun.\  {\bf 180}, 1614 (2009)
  [arXiv:0806.4194 [hep-ph]].


\bibitem{Degrande:2011ua} 
  C.~Degrande, C.~Duhr, B.~Fuks, D.~Grellscheid, O.~Mattelaer and T.~Reiter,
  arXiv:1108.2040 [hep-ph].

\bibitem{deAquino:2011ub} 
  P.~de Aquino, W.~Link, F.~Maltoni, O.~Mattelaer and T.~Stelzer,
  arXiv:1108.2041 [hep-ph].

\bibitem{Maltoni:2002qb} 
  F.~Maltoni and T.~Stelzer,
  JHEP {\bf 0302}, 027 (2003)
  [hep-ph/0208156].

\bibitem{Sjostrand:2003wg} 
  T.~Sjostrand, L.~Lonnblad, S.~Mrenna and P.~Z.~Skands,
  hep-ph/0308153.

\bibitem{Ovyn:2009tx} 
  S.~Ovyn, X.~Rouby and V.~Lemaitre,
  arXiv:0903.2225 [hep-ph].

\bibitem{ExRootAnalysis}
http://madgraph.hep.uiuc.edu/Downloads/ExRootAnalysis

\bibitem{root} 
R. Brun and F. Rademakers, Nucl. Instrum. Meth. A
389 (1997) 81.

\bibitem{Liu:2012rv} 
  S.~Liu, Y.~Mao, Y.~Ban, P.~Govoni, Q.~Li, C.~Asawatangtrakuldee and Z.~Xu,
   Phys.\ Rev.\ D {\bf 86}, 074010 (2012)
  arXiv:1205.2875 [hep-ph].

\bibitem{Pumplin:2002vw} 
  J.~Pumplin, D.~R.~Stump, J.~Huston, H.~L.~Lai, P.~M.~Nadolsky and W.~K.~Tung,
  JHEP {\bf 0207}, 012 (2002)
  [hep-ph/0201195].

\bibitem{stat:atlas} 
  The ATLAS Collaboration, CERN-OPEN-2008-020.

\bibitem{stat:2009} 
  J.~-G.~Bian, G.~-M.~Chen, M.~-S.~Chen, Z.~-H.~Li, S.~Liang, X.~-W.~Meng, Y.~-H.~Qi and Z.~-C.~Tang {\it et al.},
  Nucl.\ Phys.\ B {\bf 819}, 201 (2009)
  [arXiv:0905.2336 [hep-ex]].

\bibitem{Kwwa:2009} 
  G.~Bozzi, F.~Campanario, V.~Hankele and D.~Zeppenfeld,
  Phys.\ Rev.\ D {\bf 81}, 094030 (2010)
  [arXiv:0911.0438 [hep-ph]].

\bibitem{mcfm} 
  John M. Campbell, R. Keith Ellis, Ciaran Williams,
  ``MCFM v6.3: A Monte Carlo for FeMtobarn processes at Hadron Colliders,''
  http://mcfm.fnal.gov/

\bibitem{Kza:1997}
  U.~Baur, T.~Han and J.~Ohnemus,
  Phys.\ Rev.\ D {\bf 57}, 2823 (1998)
  [hep-ph/9710416].

\bibitem{Kwza:2010}
  G.~Bozzi, F.~Campanario, M.~Rauch, H.~Rzehak and D.~Zeppenfeld,
  Phys.\ Lett.\ B {\bf 696}, 380 (2011)
  [arXiv:1011.2206 [hep-ph]].

\bibitem{Kttbara:2011}
  K.~Melnikov, M.~Schulze and A.~Scharf,
  Phys.\ Rev.\ D {\bf 83}, 074013 (2011)
  [arXiv:1102.1967 [hep-ph]].


\bibitem{waa:2011}
  G.~Bozzi, F.~Campanario, M.~Rauch and D.~Zeppenfeld,
  [arXiv:1103.4613v1 [hep-ph]].

\end{thebibliography}
\end{document}